\documentclass[pre,superscriptaddress,showpacs,twocolumn]{revtex4}
\usepackage{epsfig,amsmath,amssymb,graphics,color,calc}
\newcommand{\be}{\begin{equation}}
\newcommand{\ee}{\end{equation}}
\renewcommand{\phi}{\varphi}
\newcommand{\ba}{\begin{eqnarray}}
\newcommand{\ea}{\end{eqnarray}}
\newcommand{\einf}{E_{\infty}}

\begin{document} 
\title{A critical test of the mode-coupling theory 
of the glass transition}

\author{Ludovic Berthier}
\affiliation{Laboratoire des Collo{\"\i}des, Verres
 et Nanomat{\'e}riaux, Universit{\'e} Montpellier II and CNRS,
34095 Montpellier, France}

\author{Gilles Tarjus} 
\affiliation{LPTMC, CNRS-UMR7600, Universit\'e Pierre et Marie 
Curie, bo\^{\i}te 121, 4 Place Jussieu, 75252 Paris cedex 05, France} 

\date{\today}
\begin{abstract}
The mode-coupling theory of the glass transition predicts the time 
evolution of the intermediate scattering 
functions in viscous liquids on the sole basis 
of the structural information encoded in two-point density
correlations. We provide a critical test of this property 
and show that the theory fails to describe the 
qualitatively distinct dynamical behavior obtained in two model liquids  
characterized by very similar pair correlation functions.
Because we use `exact' static information provided by numerical 
simulations, our results are a direct proof that some important information 
about the dynamics of viscous liquids is not captured 
by pair correlations, and is thus not described by the mode-coupling theory, 
even in the temperature regime where the theory is usually applied. 
\end{abstract}



\maketitle

\section{Introduction}

Near the glass transition, the viscosity of 
a supercooled liquid is very sensitive to small temperature changes, 
and structural relaxation is likely a thermally activated process
over an effective free-energy barrier, $E$, which grows
upon cooling, $\eta (T) \sim \eta_\infty \exp( E(T) / k_B T)$. 
From a theoretical perspective, one wishes to understand and provide
accurate predictions for $E(T)$ in terms of the microscopic 
interactions between particles in the liquid. However, even for
the simple  case of spherical particles with short-ranged, 
pairwise interactions, establishing
a link between microscopic interactions and structural relaxation
remains an open theoretical challenge~\cite{stillinger}.

In this article, we assess one of the theoretical approaches of the 
glass transition, the mode-coupling theory (MCT). 
The theory emerged in the mid-80s in the context of  the theory of 
simple liquids and  provides predictions for the time evolution of 
density autocorrelation functions from the  knowledge of the interaction 
between the particles~\cite{gotze}. Actually, the latter enters the 
theory only through two-point static density 
correlation functions. MCT therefore works under the hypothesis 
that knowledge of the pair correlation function $g(r)$ 
of the fluid (or its Fourier transform $S(q)$, the static 
structure factor)  
is sufficient to predict the dynamical evolution. We focus here on 
this hypothesis and do not address the (known) failure of MCT to 
describe activated processes in deeply supercooled liquids.

It is obvious that the evolution of the pair correlation
function has some influence on the dynamics. Accordingly, some
of the predictions that are obtained within MCT 
compare reasonably well to numerical or experimental work. 
For instance, MCT provides 
guiding information in those
cases where the changes in the static pair correlation functions brought 
about by adding either 
very short-range attractions~\cite{pham}, some 
asphericity~\cite{latz} or softness~\cite{hugo} 
to a dense system of hard spheres appear mostly responsible for the changes 
in dynamics reported for these systems.

There are however also indications in the literature that 
there may be relevant structural information which is not
captured by two-point functions. Broadly speaking, it is often 
stated that some form of  `local order' grows upon decreasing 
the temperature in supercooled liquids~\cite{stillinger}. This order can 
be for instance (frustrated) icosahedral order~\cite{tarjus} 
or the less well characterized `amorphous order' encoded in 
 high-order `point-to-set' correlation functions~\cite{pointtoset}. 
In both cases, 
$g(r)$ is essentially blind to this growing static order, and the effect 
of the latter 
on the dynamics (if any) would be completely missed 
by MCT. Similarly, it is by now understood that 
structural relaxation near the glass transition occurs in 
a spatially correlated manner, with a dynamic lengthscale 
that grows upon cooling, with again no corresponding change
in the pair correlation function~\cite{ediger}.

Recently, we found that two model liquids 
with very similar structure at the level of pair correlation 
functions display distinct behavior upon cooling~\cite{ludogilles}.
More specifically, we studied a binary Lennard-Jones (LJ) mixture~\cite{KA} and
the corresponding repulsive `WCA mixture' in which attractive forces are 
truncated beyond the minimum of the pair potential~\cite{wca}, 
and we showed that the presence or absence of the attractive forces 
determines to a great extent
 the apparent energy barrier to structural 
relaxation and its evolution with temperature and density. 
Therefore, these two model systems represent a benchmark on which the 
above ideas can be tested in much detail. 

In a previous work, Voigtmann  used liquid-state theory
to obtain the pair correlation function of the Lenard-Jones and the associated 
repulsive WCA models  in the
monoatomic case~\cite{voigtmann}. He then introduced the results 
in the MCT equations and found that the two liquids 
were essentially sharing the same dynamics, with nearly identical
MCT critical temperatures.
In our simulations on the other hand, we use a binary mixture, and 
we detect small differences between the structures of the two models. Thus, 
we directly insert the `exact' (numerically determined) structure factors 
into the MCT equations to test with a greater sensitivity 
the response predicted by MCT to small structural changes.
Our main conclusion is that MCT fails to describe the difference in the 
dynamical behavior of the two liquids. 
More generally, this conclusion also suggests that microscopic 
theories of the glass transition based on pair correlation functions
cannot accurately predict the evolution of the structural 
relaxation time. 

In Sec.~\ref{model} we describe our models and method of analysis.
In Sec.~\ref{dynamics} we present our results for the dynamics.
Section \ref{conclusion} concludes the paper.

\section{Models, methods and static results}

\label{model}

\subsection{Static structure}

We consider the binary mixture of Lennard-Jones particles
introduced by Kob and Andersen~\cite{KA}, as well as its WCA
truncation~\cite{wca}. The system is a 80:20 mixture 
of $A$:$B$ particles interacting with 
the following interatomic pair potential between species $\alpha$ and 
$\beta$, with $\alpha, \beta = A, B$:
\begin{equation}
v_{\alpha \beta}(r) = 4\epsilon_{\alpha \beta} 
\left[\left( \frac{\sigma_{\alpha \beta}}{r}\right)^{12}- \left( 
\frac{\sigma_{\alpha \beta}}{r}\right)^{6} + C_{\alpha \beta} \right], 
\; r \leq 
r_{\alpha \beta}^c,
\label{potential}
\end{equation}
and $v_{\alpha \beta}(r > r_{\alpha \beta}^c)=0$.
In this expression, the cutoff distance $r_{\alpha \beta}^c$ is equal to the 
position of the minimum of $v_{\alpha \beta}(r)$ for the WCA potential and 
to a conventional cutoff of $2.5 \sigma_{\alpha \beta}$ for the standard LJ 
model; $C_{\alpha \beta}$ is a constant such that 
$v_{\alpha \beta}(r_{\alpha \beta}^c) =0$. The value of the
parameters $\sigma_{\alpha \beta}$ and $\epsilon_{\alpha \beta}$
were designed to obtain good glass-forming ability~\cite{KA}.

We perform molecular dynamics simulations of both systems 
in the $NVE$ ensemble after proper equilibration at 
a chosen state point characterized by a density $\rho$ and a temperature $T$. 
We use $N=900-1300$ particles (depending on the
density) and work with
periodic boundary conditions. A broad range of density has 
been considered with $\rho$ varying from $1.2$ to $1.6$. Lengths, temperatures 
and times are given in 
units of $\sigma_{AA}$, $\epsilon_{AA}/k_B$, and $(m \sigma_{AA}^2/ 48 
\epsilon_{AA})^{1/2}$ respectively. In line with the WCA theory~\cite{wca}, 
the two liquid models are compared at the same $(\rho, T)$ state points. 
Their pressures then differ, with the attractive interaction roughly 
providing a temperature independent negative shift.

\begin{figure}
\psfig{file=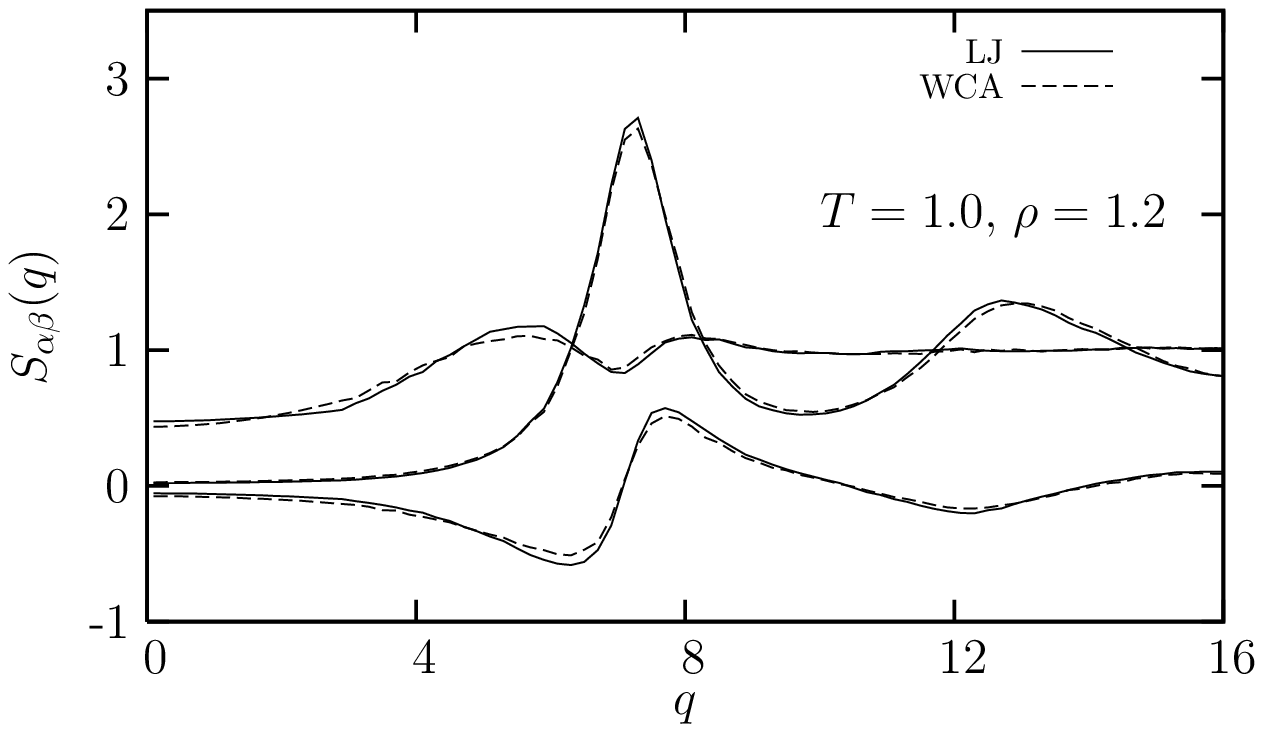,width=8.5cm}
\psfig{file=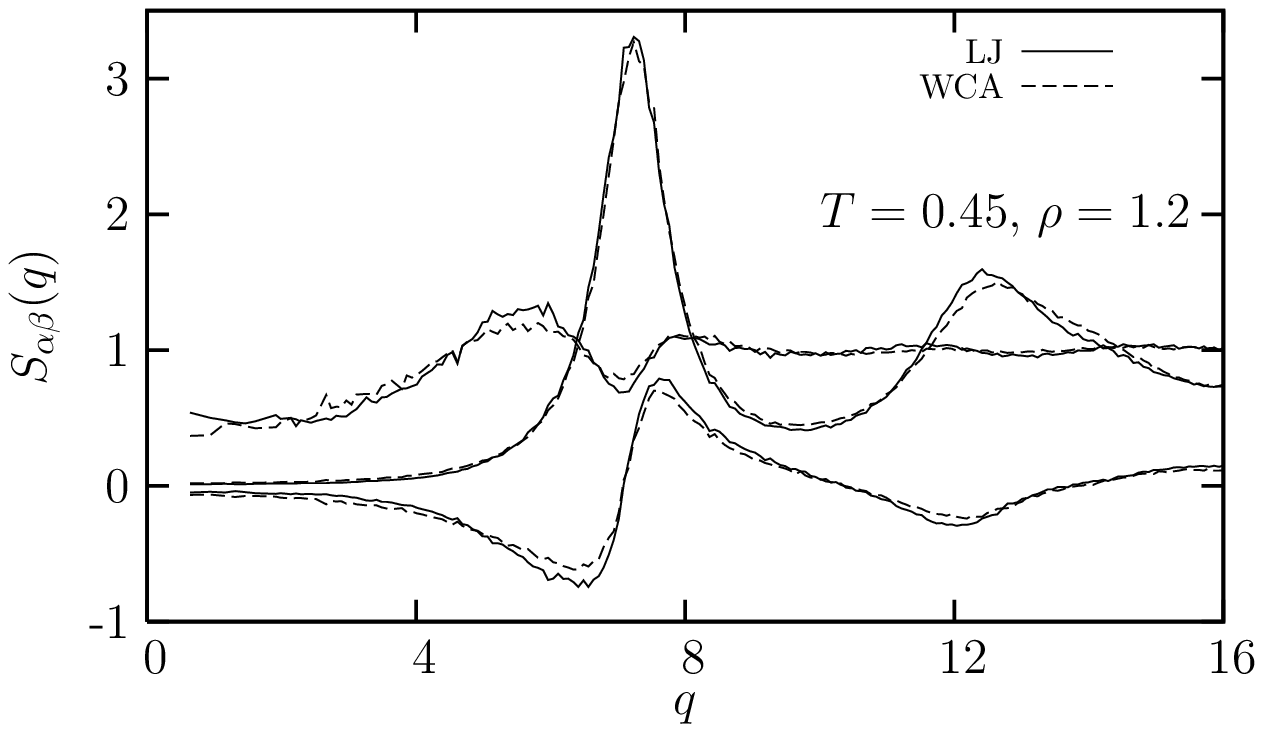,width=8.5cm}
\caption{\label{sq12}
Partial static structure factors $S_{BB}(q)$, $S_{AA}(q)$, 
and $S_{AB}(q)$ (from top to bottom at small $q$)
at $\rho=1.2$ near the 
theoretically determined $T_c$ of MCT (top) and close to the 
$T_c$ obtained from fitting the simulation data (bottom). 
The pair structures of the  
LJ and WCA liquids are very close, and the small deviations 
seem to increase somewhat when $T$ decreases.}
\end{figure}

From the numerical simulations we obtain both static and dynamic
properties. We use static information as input for the
MCT analysis described below, while the dynamical data are 
used to test the accuracy of the 
theoretical results. 
For the dynamics, we record the evolution of the time 
dependence of the self part of the 
intermediate scattering function 
\begin{equation}
F_s^\alpha(q,t)=\frac{1}{N_\alpha} \left\langle \sum_{j=1}^{N_\alpha} 
e^{i \mathbf{q}.
(\mathbf{r}^\alpha_j(t)-\mathbf{r}^\alpha_j(0))} \right\rangle,
\end{equation}
with $q \sigma_{AA} \simeq 7.2$,
which corresponds to the position of 
the peak of the total static structure factor at $\rho=1.2$, 
see Fig.~\ref{sq12}. 
In this expression, $N_\alpha$ denotes the number of particles
of species $\alpha$, and ${\bf r}_j^\alpha(t)$ is the position
of particle $j$ belonging to species $\alpha$ at time $t$.
From the decay of $F_s^\alpha(q,t)$ for the majority species $\alpha=A$, 
we obtain the relaxation time $\tau(\rho,T)$
which we conventionally define as $F_s^A(q,\tau)=1/e$.
Some of the data for $\tau(\rho,T)$ were presented in 
Ref.~\cite{ludogilles}.

At the structural level, we measure the partial 
structure factors $S_{\alpha \beta}(q)$ which are needed
as input for the MCT calculations. They are defined as
\be
S_{\alpha \beta}(q) = \frac{1}{\sqrt{N_\alpha N_\beta}} 
\sum_{m=1}^{N_\alpha} \sum_{n=1}^{N_\beta} e^{i \mathbf{q}.
(\mathbf{r}^\alpha_m-\mathbf{r}^\beta_n)},
\label{Sab}
\ee 
which can be written in a more compact form as a 
$2 \times 2$ matrix ${\bf S}(q)$ whose matrix
elements are the partial structure factors 
$S_{\alpha \beta}(q)$. 

\begin{figure}
\psfig{file=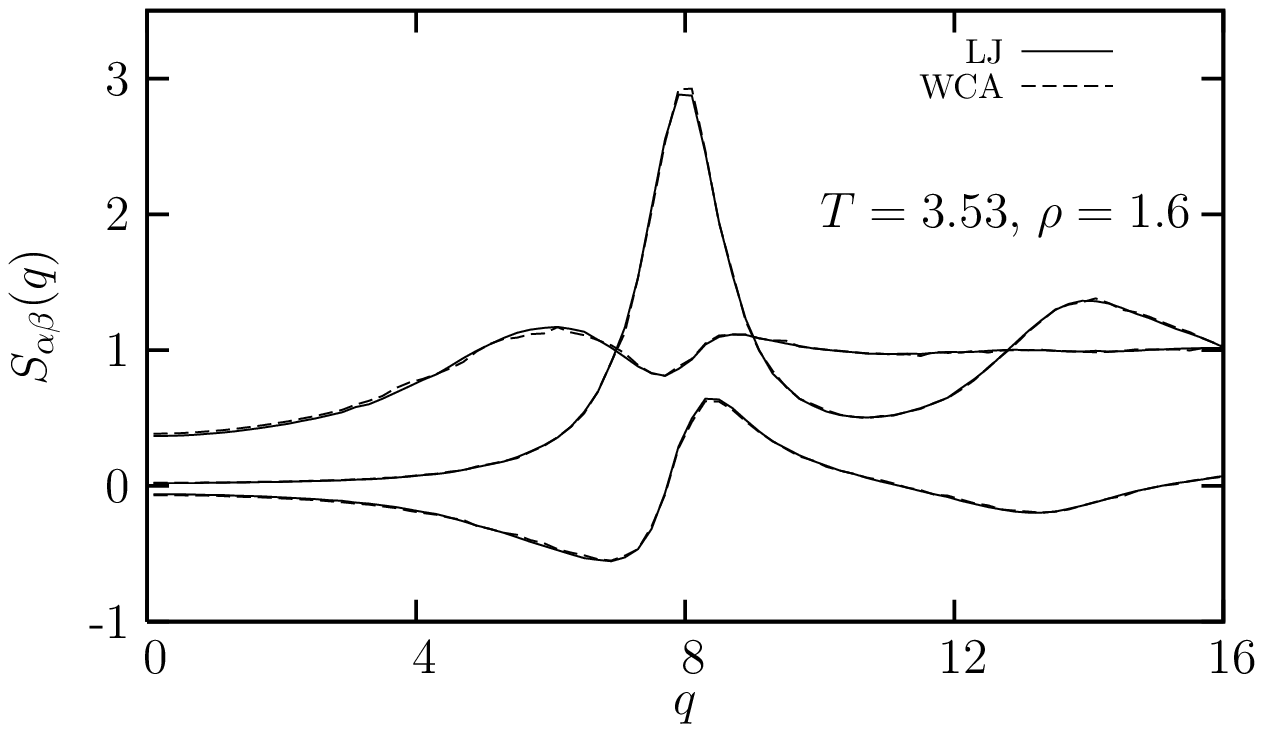,width=8.5cm}
\psfig{file=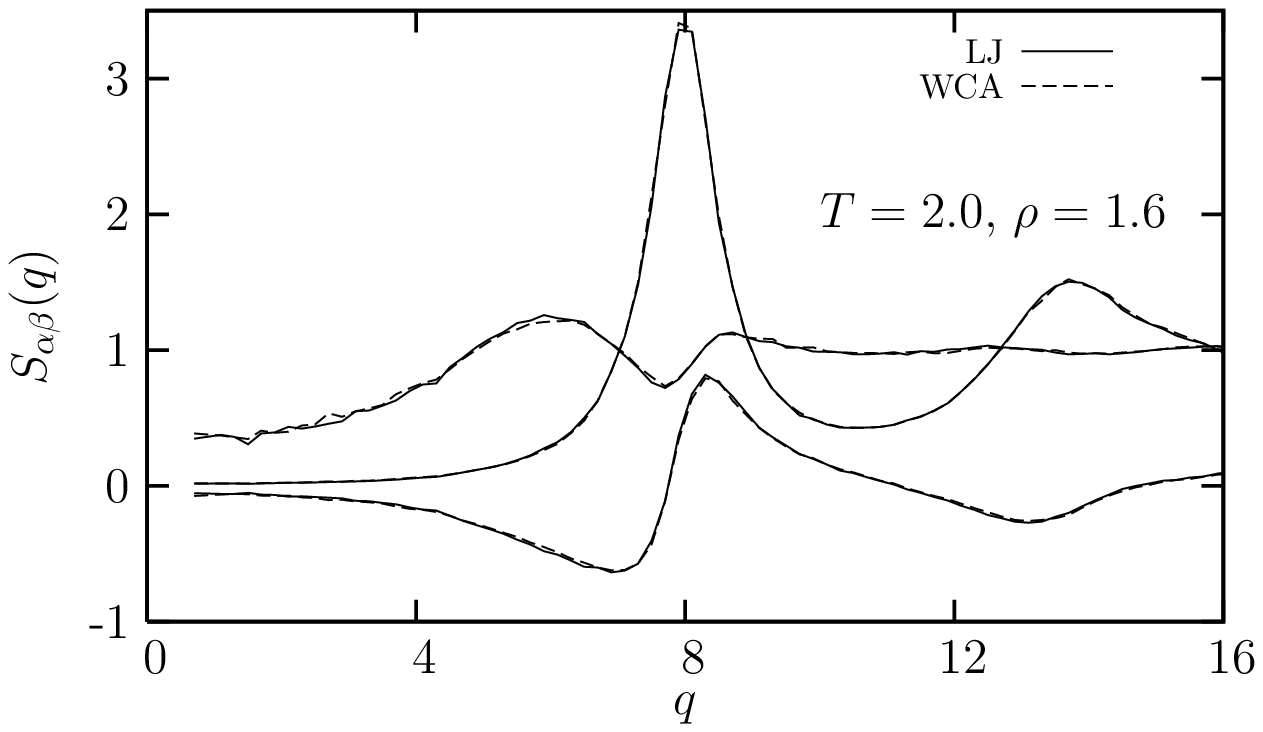,width=8.5cm}
\caption{\label{sq16}
Same as Fig.~\ref{sq12} for $\rho=1.6$. Here the agreement 
between LJ and WCA mixtures is nearly perfect at all temperatures.}
\end{figure}
 
We present a set of representative results for the partial structure
factors in Fig.~\ref{sq12} for $\rho=1.2$ and in Fig.~\ref{sq16}
for $\rho=1.6$. The former corresponds to the canonical density 
at which the Kob-Andersen mixture is usually studied~\cite{KA}.
For each density, we present the data at two temperatures, 
corresponding to the  
location of the MCT transition determined either theoretically or 
by fitting the simulation data. 
Generically, the latter is lower than the former. 

These data confirm the well-known fact that 
the LJ and WCA models have a pair structure which is very similar, even 
in the case of a binary mixture. This supports the idea that
the contribution of the attractive forces at the level 
of the two-point structure can be considered as a small perturbation
and treated as such in theoretical liquid-state calculations~\cite{wca}.
Looking more closely, we can nevertheless detect 
some small differences between the two systems. These 
differences increase slightly when one lowers the temperatures 
at constant density, as can be concluded for instance by comparing the 
two panels in Fig.~\ref{sq12}. The agreement between
the pair structures of the two liquids becomes excellent when 
density increases, 
see Fig.~\ref{sq16}. This is physically reasonable since 
the steep repulsive core of the pair potential in Eq.~(\ref{potential}), 
which is the same
for both models, should play a predominant role under those thermodynamic
conditions.

We have obtained ${\bf S}(q)$ for a large number of state 
points at the densities $\rho=1.2$, 1.4 and 1.6. We have then proceeded
in two steps. First, we have directly inserted 
those data into the MCT dynamical equations presented below
to obtain a rough determination of the location of the
MCT transition.
Then, we have used a smooth interpolation procedure 
to obtain the partial structure factors in a small temperature window
near the transition because it is not possible to resolve
the structure directly from simulations with the accuracy needed
to characterize the MCT transition. This interpolation 
procedure is however quite innocuous, as the structure evolves 
in a mild and continuous manner 
in the temperature regime where the interpolation is performed.
 
\subsection{Mode-coupling theory}

The mode-coupling theory of the glass transition~\cite{gotze} 
was originally derived to describe the dynamics of Newtonian 
systems~\cite{bengtzelius}. (A somewhat different derivation was 
also provided in the framework of the  nonlinear fluctuating 
hydrodynamics~\cite{Das-Mazenko}.) An analogous theory was later 
derived for Brownian systems~\cite{sl}. Here, we briefly present 
the latter version of MCT which, for convenience, we use in the following.

The starting point of the theory is an exact equation for the time 
evolution of the intermediate scattering functions ${\bf F}(q;t)$ in terms
of the so-called irreducible memory function,
\begin{align}
\label{FMirr}
& & 
\frac{\partial}{\partial t} 
 {\bf F}(q,t)  =  - D_0 q^2 {\bf S}^{-1}(q) {\bf F}(q,t) 
\nonumber \\ 
& & - 
\int_0^t dt' {\bf M} (q,t-t') \frac{\partial}{\partial t'}  {\bf F}(q;t').
\end{align}
In this equation, $D_0$ is the diffusion coefficient of an 
isolated Brownian particle, 
and ${\bf F}(q,t)$ is a matrix whose elements are the 
intermediate scattering functions $F^{\alpha \beta}(q,t)$
defined as
\be
F^{\alpha \beta}(q,t) =  \frac{1}{\sqrt{N_\alpha N_\beta}} 
\sum_{m=1}^{N_\alpha} \sum_{n=1}^{N_\beta} e^{i \mathbf{q}.
(\mathbf{r}^\alpha_m(t)-\mathbf{r}^\beta_n(0))},
\label{Fab}
\ee
and is such that ${\bf F}(q,t=0)={\bf S}(q)$.
The matrix of memory functions, ${\bf M}(q,t)$, 
can be expressed, in the mode-coupling approximation, 
in terms of the intermediate scattering functions ${\bf F}$ and 
structure factors ${\bf S}$. The corresponding 
(lengthy) expression can be found in 
Ref.~\cite{flennerszamel}.

Equation (\ref{FMirr}) allows us to evaluate the time 
dependence of the
intermediate scattering functions. The only input required is the 
static structure factor ${\bf S}(q)$. It is 
easy to see that 
the natural time unit for our system is $\sigma^2/D_0$.
In the following, all times are given in terms of this unit. 

Numerical resolution of Eq.~(\ref{FMirr}) is difficult because
one needs to describe the evolution of the intermediate scattering function on
very widely separated time scales. 
The commonly used algorithm
was first described in Ref.~\cite{fuchs}; here, we use the 
implementation described 
in great detail in Ref. \cite{flennerszamel}. Briefly, 
the basic steps of the algorithm are as follows. The integro-differential 
equation is discretized and solved numerically for $2 N_s$ steps with
a finite time step of $\delta t$.
After $2 N_s$ steps are completed, the time step is doubled and the
results from the initial $2 N_s$ steps are mapped onto a new equally
spaced set of $N_s$ values for the quantities needed to continue the
numerical algorithm. This mapping includes the integrals as well as 
the intermediate scattering functions. Then,
the numerical algorithm is restarted with the new time step and
proceeded for another $N_s$ time steps, and the mapping is performed 
again. This procedure is followed until a convergence condition is satisfied.
In the present work we used 200 equally spaced wave-vectors with spacing 
$\delta = 0.2$, the first wave-vector being at $k_{\rm min}=\delta/2$, and the
largest wavevector at $k_{\mathrm{max}} = 39.9$. 

\section{Dynamical behavior}

\label{dynamics}

\subsection{Intermediate scattering functions}

\begin{figure}
\psfig{file=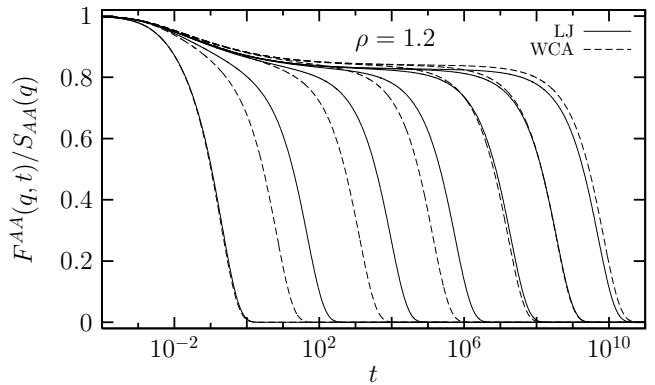,width=8.5cm}
\caption{\label{fq}
Normalized intermediate scattering function for the majority 
component for $\rho=1.2$ 
and $q = 7.1$, 
for both LJ (full lines) and WCA (dashed lines) models
obtained from the numerical solution of Eq.~(\ref{FMirr}).   
The temperature decreases from left to right. 
LJ: $T=3.0$, 1.0, 0.908, 0.899, 0.8975, 0.8972, 0.8971.
WCA: $T=3.0$, 1.0, 0.77, 0.746, 0.7425, 0.7421, 0.74198.
Note that the two sets of temperatures are not identical. The 
lower temperatures in each set provide a good estimate 
of the $T_c$ values for each model.}
\end{figure}

The solution of the MCT dynamical
equations provides the time evolution of the intermediate scattering 
functions $F^{\alpha \beta}(q,t)$. We present a selection of  
data for the majority component, $\alpha=\beta=A$, at density $\rho=1.2$ in 
Fig.~\ref{fq}. These data have the standard shape obtained 
within MCT, whereby the time decay streches over many 
decades of time when approaching the MCT singularity
and occurs in two widely separated timescales. 
From now on, we focus on the slower of these, corresponding to 
the structural (alpha) relaxation of the system.
Its time dependence is empirically well described 
by a stretched exponential form, 
\be
F(q,t) \sim \exp [ - (t/\tau)^\beta ], 
\label{stretched}
\ee
as commonly found in supercooled liquids. 
The data in Fig.~\ref{fq} suggest that 
the stretching exponent $\beta$ is essentially the same 
at all temperatures (and so MCT data obey a time-temperature
superposition property) and is very close for both 
LJ and WCA liquids, with $\beta \approx 0.87$.
We find similar values and agreement 
for all densities up to $\rho=1.6$. 
For $\rho=1.2$ and $q=7.25$, Kob and Andersen report the value
$\beta = 0.85$~\cite{KA2}. 

The intermediate scattering functions of the two models superimpose 
at very high temperatures, $T=3.0$, when the structure factors 
also show very good agreement. However, when temperature is 
decreased, the dynamics of the WCA model becomes faster than 
that of the LJ model, see the data for $T=1.0$ in Fig.~\ref{fq}. 
The difference becomes more dramatic as the mode-coupling singularity 
is approached and 
long relaxation times are obtained. 
Although this is reminiscent of our numerical findings on the role 
of attractive forces in supercooled liquids~\cite{ludogilles},
we shall see below that MCT only marginally accounts
for this effect. 

As in numerical simulations, we determine the structural
relaxation time $\tau(T,\rho)$ through the 
relaxation $F^{\alpha \beta}(q,\tau) / S_{\alpha \beta}(q) = 1/e$. 
This relaxation time depends on the wavevector $q$, 
the chosen species $\alpha$ and $\beta$. It is \textit{a priori} different
from the time extracted from simulations where 
only the incoherent part of the intermediate  
scattering function is considered. However, a simplifying feature 
of MCT calculations is that the evolution of 
$\tau$ is the same for all observables near the 
transition. Thus, it does make sense to focus 
on a given wavevector (we choose the maximum of the 
first diffraction peak), to focus on the majority component
of the liquid ($\alpha = \beta = A$), and to compare
these results to the simulation data for which the incoherent
scattering function is measured. 

The only difference between the two sets of data shown
in Fig.~\ref{fq} lies in the temperature regimes they cover
since the MCT transition of the WCA model occurs at a slightly
lower temperature than that of the full LJ model including 
the attractive component of the potential. We now turn to 
a more precise discussion of this difference. 

\subsection{Relaxation times}

We now enter the core of our analysis and study 
in detail the evolution with $T$ and $\rho$ of the structural
relaxation time, as predicted by MCT, and compare these theoretical
results to those from computer simulations.

\begin{figure}
\psfig{file=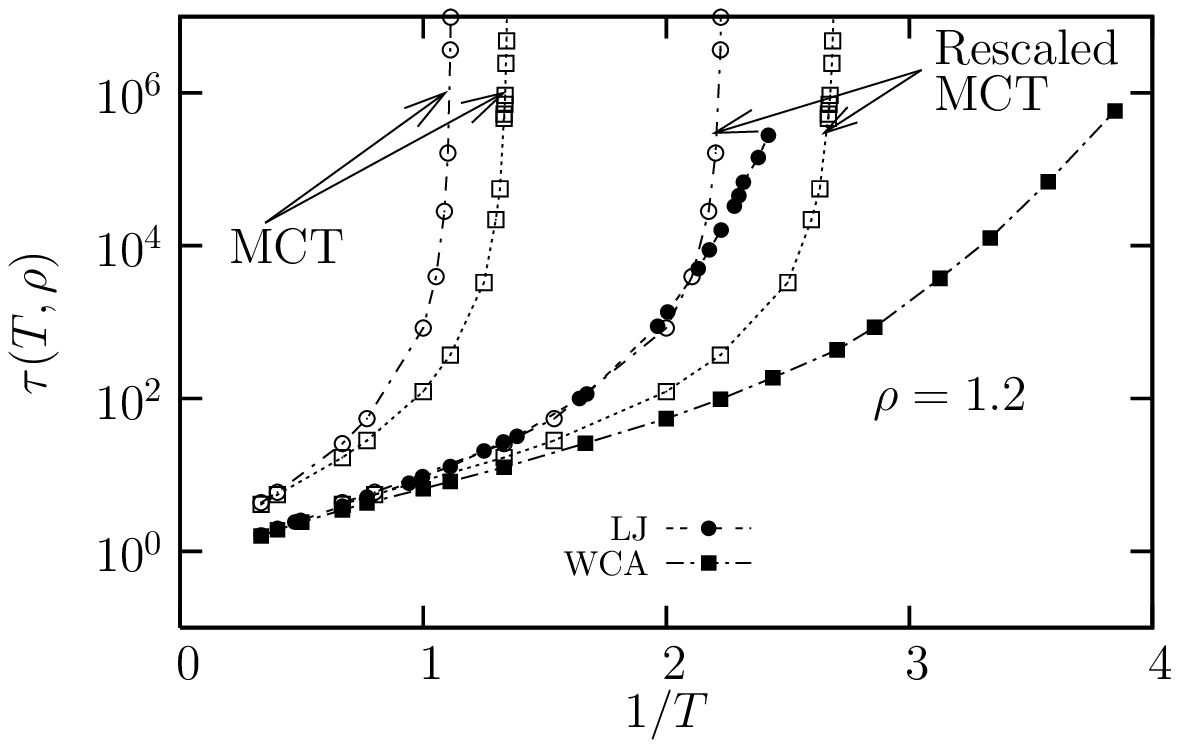,width=8.5cm}
\psfig{file=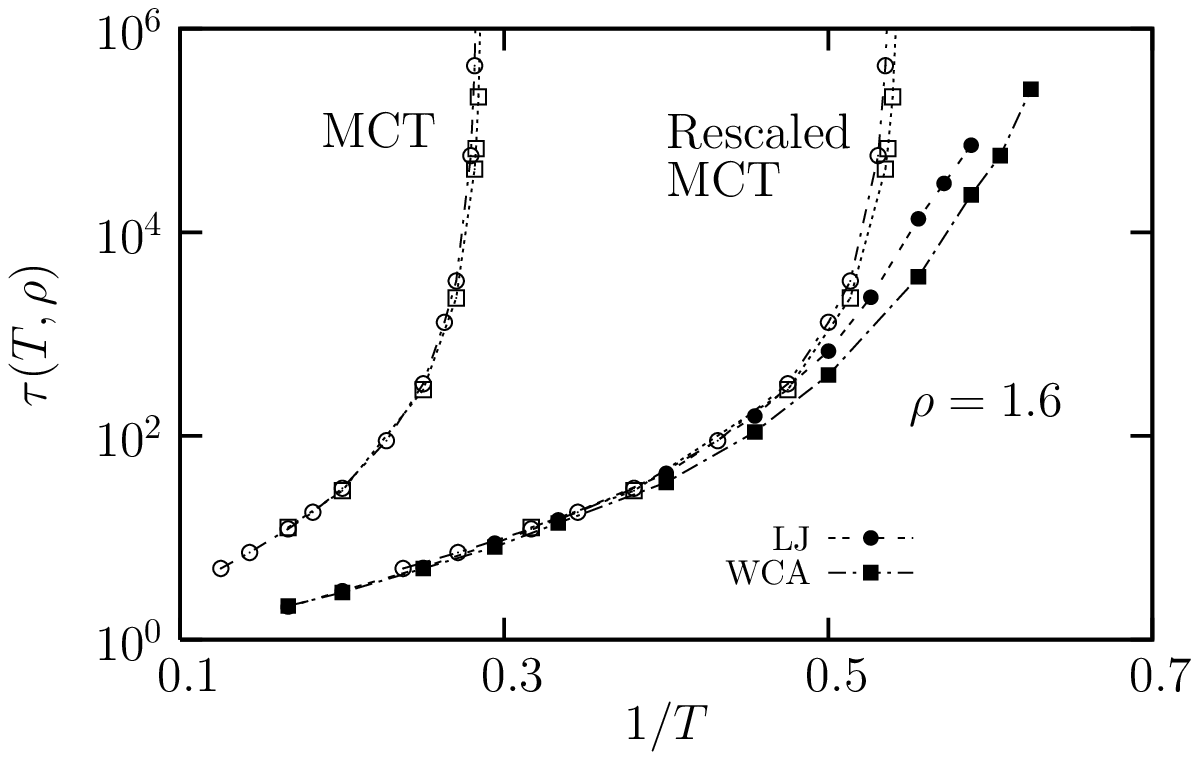,width=8.5cm}
\caption{\label{comp}
Comparison of the relaxation times obtained by MCT (open symbols) and 
simulations (filled symbols) 
at $\rho=1.2$ (upper panel) and $\rho=1.6$ (lower panel). 
We both show the bare MCT data and the data rescaled to provide 
the best collapse of the LJ data. The rescaling factor for the temperature is 
2 for $\rho=1.2$ and 1.9 for $\rho=1.6$. 
This same factor is then applied with no further changes to the WCA 
model.}
\end{figure}

We first present in Fig.~\ref{comp} our results in an Arrhenius
plot, showing the evolution of $\tau(T,\rho)$ on a log-scale 
against the inverse temperature $1/T$, for both the LJ and the WCA models, 
as measured in MCT calculations and in simulations. 
We show results for $\rho=1.2$ and $\rho=1.6$. 

In this Arrhenius representation, the results of MCT
calculations appear almost `vertical', because MCT predicts
the existence of a critical singularity $T_c$ at which the 
relaxation time diverges algebraically~\cite{gotze}.
This divergence is not seen in computer simulations and so the
simulation data look qualitatively different. 

A second striking observation from Fig.~\ref{comp} 
is that MCT clearly overestimates the temperature 
regime where slow dynamics sets in by about 100~\%. Thus, 
there is no range of temperature where the theoretical calculations 
follow the simulation results, even if a vertical
shift (corresponding to a trivial rescaling of the microscopic
timescale) is allowed. 

At high temperatures, a good empirical 
representation of both MCT and simulation results is 
obtained by introducing an Arrhenius-like temperature dependence, 
\be
\tau \sim \tau_\infty \exp \left( \frac{ \einf }{T} \right).
\label{arrhenius}
\ee
Although there is no specific theoretical foundation 
for such a fit, it often accounts quite well for high
temperature results, even in experiments~\cite{gilles4}. 
When fitting the data to Eq.~(7), we obtain the values 
$\einf \approx 5.5$ and 4.2 for the LJ and WCA systems within MCT, 
while the simulation data are characterized by $\einf \approx 2.55$ and 2.0.
The different values found for the two models in the high-temperature
regime reflect the fact that attractive forces 
already play an important quantitative role in this regime,
despite the very little changes observed in static quantities.
This effect was reported before in the case of simple 
liquids~\cite{berne,kushick,szamel}. 

The role of attractive forces at high temperatures
is partially captured by our MCT calculations, 
since we find theoretically a 25~\% change in $\einf$ between
the two models. However, given that the predicted values are 
off by a factor of $2$, this agreement could well be fortuitous.
At a density $\rho=1.6$, we find $\einf \approx 12$ from simulations of the two
systems, while MCT again overestimates this value and predicts
$\einf \approx 22.5$ for both models. So, at high temperatures 
and large densities, the difference between LJ and WCA mixtures 
 vanishes both in simulations and MCT calculations, 
but the activation energy predicted by the latter is still in error by a 
factor close to 2. 
It is interesting to note that there is nearly the same factor 2
in error for $\einf$ and for $T_c$ (see below), 
which suggests that MCT does not merely `break down' 
at low $T$ but is in fact always quite inaccurate
in its prediction for the temperature evolution 
of the structural relaxation time, even in the high-temperature liquid. 

The strong disagreement between MCT results and simulations 
is well-known~\cite{gotze} and is two-fold: (i) MCT overstimates
the temperature regime where slow dynamics takes place; (ii)
the predicted algebraic divergence of the
relaxation time is not really observed. When fitting data to MCT 
predictions, one can thus only find a modest temperature
regime where an algebraic divergence describes the 
data and one must simultaneously resort to 
some sort of rescaling or shift of the control parameters. 
Since this gives considerable freedom in the data treatment, 
we adopt the following procedure, shown in Fig.~\ref{comp}. 
At a given density, we first concentrate on the LJ model. We then rescale 
$\tau$
and $T$ by adjustable time and temperature scales 
in order to obtain the `best' collapse of theory
and simulations. We obtain a temperature scaling factor of
2 for $\rho=1.2$, and 1.9 for $\rho=1.6$, which
mirrors the factor 2 in $\einf$ found above at high temperatures. 
Once this rescaling is performed for the LJ model, we use the same 
scaling factor for temperature in the WCA model. The results are presented 
in Fig.~\ref{comp}. While the rescaling works well if $T$ is not 
too low in the LJ system, it performs very poorly for WCA, even when 
density is as large as $\rho=1.6$. 
This implies that the different dynamical behaviors 
observed in simulations for LJ and WCA models~\cite{ludogilles} is  
only very qualitatively reproduced by MCT
 and is considerably underestimated  by the theory at a quantitative level. 

\begin{figure}
\psfig{file=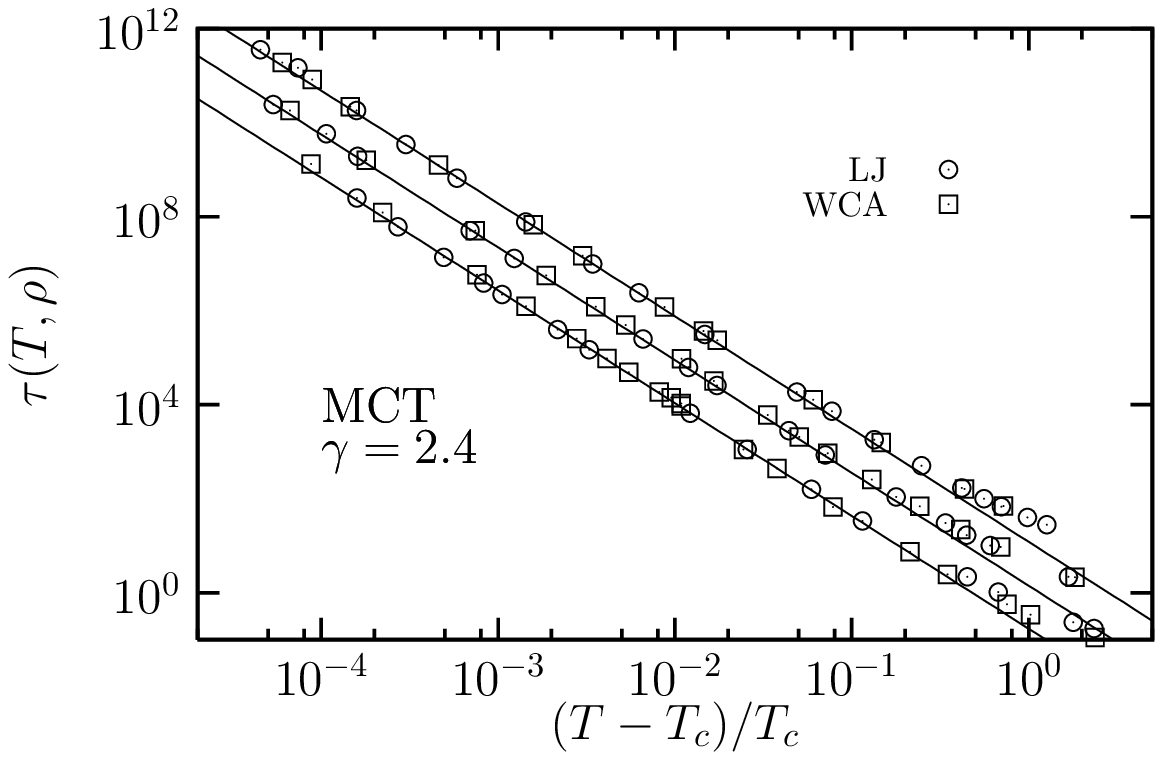,width=8.5cm}
\psfig{file=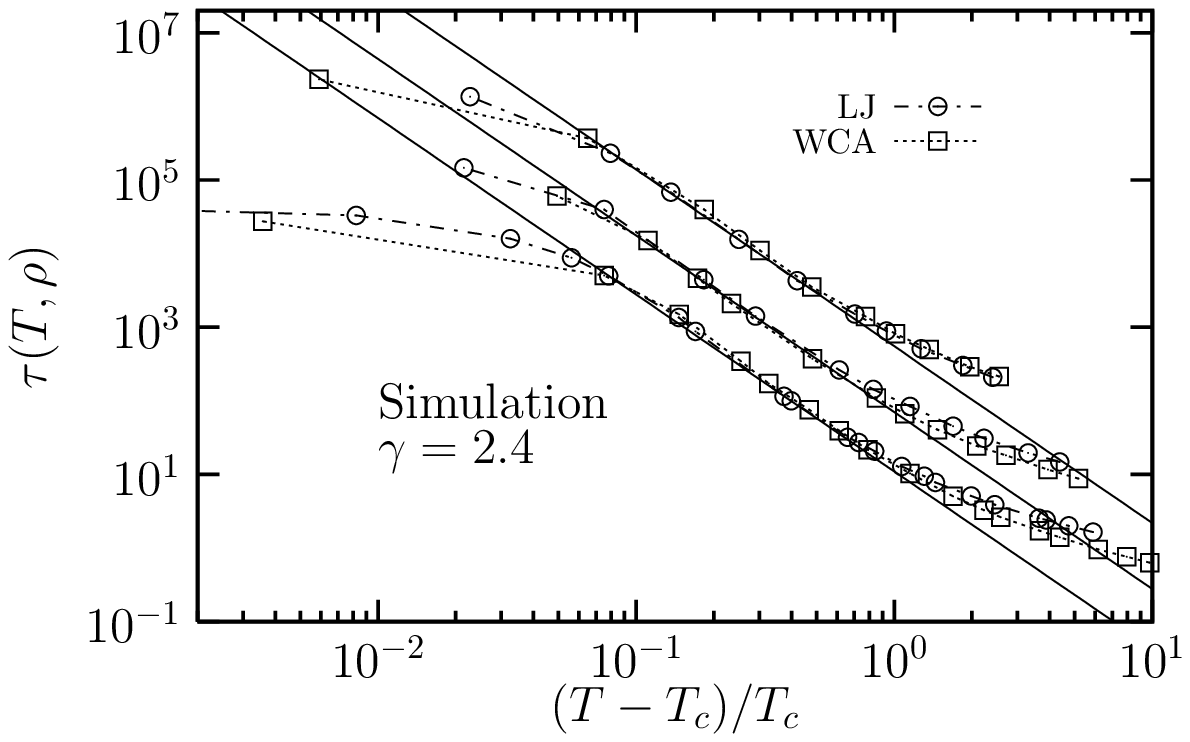,width=8.5cm}
\caption{\label{taumct}
MCT plot of the relaxation times in which 
the algebraic divergence in Eq.~(\ref{mctfit})
appears as a straight line. 
Top: theoretical curves fitted with the same $\gamma=2.4$
for $\rho=1.2$, 1.4, and 1.6 (from bottom to top).
Bottom: same for numerical simulations, but note the
difference in scales and the deviations occurring at low temperatures.
Data at different densities are vertically shifted, for clarity.}
\end{figure}

This last statement can be made more precise 
by determining the location of the MCT singularity. 
We first determine $T_c$ within the theory by following 
the growth of $\tau$ near the transition, which is well
described by an algebraic divergence:
\be
\tau \sim (T-T_c)^{-\gamma}.
\label{mctfit}
\ee
The value of the critical exponent $\gamma$ depends in
principle on the studied model. 
In Fig.~\ref{taumct}, we show however that the value $\gamma=2.4$
actually describes both the LJ and the WCA models for densities
$\rho=1.2$, 1.4 and 1.6. The values of the critical 
temperatures are reported in Table \ref{table}: $T_c$ increases
by a factor $\approx 4$ when density increases from 1.2 to 1.6.
As density increases, the difference in critical temperatures 
between the LJ and WCA models becomes smaller, decreasing from 
17~\% at $\rho=1.2$ to 1~\% at $\rho=1.6$ (within the theory).

While the determination of $T_c$ is unambiguous in MCT calculations,
this is not the case in simulations where the MCT 
power law cannot be followed for arbitrarily large relaxation times. 
In practice, this means that the values of $T_c$ and $\gamma$ obtained 
in simulations are strongly correlated and therefore depend on the chosen 
temperature range for fitting. Thus, we decided to constrain our 
data analysis
by imposing the value of the exponent $\gamma$ obtained 
in the theoretical calculations. We then determined the value of 
$T_c$ that fits the data best with this exponent. The quality of the fits can 
be judged in Fig.~\ref{taumct}, where the data for $\tau(T,\rho)$ 
are plotted as a function of the reduced variable $(T-T_c)/T_c$ in 
a log-log representation where Eq.~(\ref{mctfit}) should 
appear as a straight line.
The fit holds over about 2-3 decades in relaxation time, as
commonly found in the analysis of numerical and experimental
data~\cite{gotze}, and deviations at low temperatures 
occur similarly for both models at all densities. 
As noted before~\cite{flennerszamel}, it should be remarked that 
the algebraic fit to simulation data is obtained in a range of 
$(T-T_c)/T_c$ where the theoretical predictions have not yet entered the 
critical regime, which casts some doubts on the consistency
of the fitting procedure.

\begin{table}
\begin{tabular}{|| l || l | l | l ||}
\hline
  & 1.2  & 1.4   &  1.6   \\
\hline 
LJ - MCT         &  0.8971  & 1.8677   & 3.528   \\
LJ - Simulation  &  0.435   & 0.93     & 1.76    \\
WCA - MCT        &  0.7419  & 1.7707   & 3.489   \\
WCA - Simulation &  0.28    & 0.81     & 1.69    \\
\hline
\end{tabular}
\caption{\label{table} Theoretical critical temperatures $T_c$
for the LJ and WCA models at different densities, and comparison
with estimates from fits to the simulation data.} 
\end{table}

The values of the critical temperatures obtained by fitting the simulation 
data are also reported 
in Table \ref{table}. As found theoretically, the 
difference between the LJ and WCA models is smaller at larger 
density, when the structure factors become more similar
(see Fig.~\ref{sq16}). However, the difference in critical 
temperatures decreases from 36~\% at $\rho=1.2$ to 4~\% 
at $\rho=1.6$ and is much larger than the MCT theoretical predictions 
at all densities
(respectively 17~\% and 1~\%). 
Thus, the difference in the dynamical behavior of the two models is only 
marginally captured by MCT calculations and is 
quantitatively considerably underestimated. 

\section{Discussion}

\label{conclusion}

The structure of simple liquids is usually described in 
terms of pair correlation functions, and the development of 
analytical theories to predict their 
evolution with density or temperature
from the knowledge of the interaction between particles 
has been a major theoretical achievement~\cite{hansen}.
Whether knowledge of $g(r)$ is enough to characterize 
the structure and predict the dynamics 
of viscous liquids has however been an open question in the field 
of the glass transition. This is a central issue for 
mode-coupling theory which 
makes dynamical predictions from the sole knowledge of two-point
density correlations.

To address this issue we have used the pair correlation functions
taken from molecular dynamics simulations to obtain MCT
predictions for the relaxation time of two model liquids
characterized by similar pair structures, but very distinct 
dynamics~\cite{ludogilles}. We have found
that MCT is essentially blind to these dynamical changes, 
being unable to account for them in a quantitative way. 
In this case at least,  
the necessary structural input for the dynamics of the viscous 
liquids is not merely encoded
in two-point density correlation functions. Accordingly, MCT is bound 
to yield quantitatively inaccurate
predictions. As mentioned in the introduction, there still remains 
room for applying MCT, as there exists physically relevant 
glassy phenomena associated to important changes in the structure 
that are captured by two-body correlations.

In a similar vein, the present work has been an opportunity to test the idea, 
popularized in particular in the context of MCT, that even small changes
in the pair structure may have dramatic consequences in the 
dynamics. For the case under study, we have found that 
the small differences seen between the structure factors of the two 
liquid models 
only produce, within MCT, minor dynamical differences, thus confirming
Voigtmann's results~\cite{voigtmann}. As already stated, these 
MCT predictions dramatically underestimate
the changes observed in the simulations. An additional factor
explaining this discrepancy is the fact that MCT overestimates
by a large amount the location of the putative 
singularity (empirically estimated by fitting simulation data). As a 
result, MCT predictions are 
based on structure factors measured at relatively high 
temperatures where differences between the WCA and LJ models
are even less pronounced: compare both panels in Fig.~\ref{sq12}.  

Finally, our work does not address whether the failure exposed here of the 
`$g(r)$ determines the dynamics' MCT motto that is shown here 
is connected to the 
failure of MCT to describe activated dynamics at low temperature. 
For instance, in their attempt to incorporate 
activated processes in the MCT framework, Schweizer and coworkers~\cite{ken}
manage to avoid the MCT algebraic divergence of 
Eq.~(\ref{mctfit}). However, in their approach,
it is still the pair correlation function which determines
the amplitude of the free energy barriers that have to be crossed 
dynamically. In this case,
one also expects the theory to miss the difference of dynamical 
behavior between the
LJ and WCA liquid models, despite the presence of activated processes.

To conclude, our work shows that structural information not incorporated 
in two-body density correlations likely play an important 
quantitative part in driving the slowdown of dynamics in viscous 
liquids. This raises serious doubts on the ability 
of a number of analytical approaches to make quantitative 
predictions in supercooled liquids. In this paper, we have only focused 
on MCT  and we leave for a future publication
a more general discussion of the consequences of this finding.  
Our results also motivate further research to detect
more complicated forms of `hidden', `amorphous', or `frustrated'
order in liquids approaching the glass transition.

\acknowledgments
We wish to thank E. Flenner and G. Szamel for generously making their 
numerical code to solve MCT equations for a binary 
mixture~\cite{flennerszamel} available to us, thereby making this 
work possible.
We also thank W. Kob for useful advice.
We acknowledge partial support from the ANR Dynhet.

\end{document}